\documentstyle[preprint,aps]{revtex}
\begin{document}
\def\eps{\epsilon}
\def\tc{$T_c$\ }
\def\214{$2\,1\,4$}

\begin{title}
{Anomalous Magnetothermal Resistance of High-$\bf T_c$
Superconductors: Anomalous Cyclotron Orbits at a Dirac Point}
\end{title}

\author{Philip W. Anderson}

\address {Joseph Henry Laboratories of Physics\\
Princeton University, Princeton, NJ 08544}
\maketitle

\begin{abstract}
In trying to explain the anomalously sharp dependence of thermal
conductivity on magnetic field in the cuprate high $t_c$ 
superconductors, it was found that previous discussions of
quasiparticle motion near the gap nodes were in error because of
failure to take into account total current conservation. We
present corrected equations of motion for quasiparticle motion
and discuss their relationship to the observations. 

\end{abstract}
%\section{Introduction}
\vfill\eject

In thermal conductivity of High $T_c$ cuprates, at all
temperatures below $T_c$, there is observed a remarkably steep
dependence on applied $c$-axis magnetic fields even as low as .1T.
Fig (1) shows observations in both
$(La-Sr)_2CuO_4$ and YBCO\cite{1} From thermal Hall
(Righi-LeDuc) effect, it was concluded that the enhancement of
thermal conductivity below \tc is electronic, and that large
fields seem to block this conductivity. But one of the primary
mysteries has been the very sensitive dependence on magnetic
fields in which any Zeeman or cyclotron frequency corresponds to
an energy $<1^\circ$K, much lower than thermal energy
$\sim30$--$60^\circ$K in either case.

It has been clear that the carriers responsible for electronic
thermal conductivity at lower temperatures must be those near the nodes of the
``$d-$symmetry'' energy gap at $\pm k_F$, $\pm k_F$. Although their
density is relatively small, the very rapid removal below $T_c$,  as
the gap opens up, of the electronic scattering mechanism
responsible for the normal state ``T-linear'' resistivity could
allow  peaking of quasiparticle conductivities.\cite{2} 
The ARPES data on electron Green's functions near the gap nodes,
however, do not support this interpretation, and it seems more
plausible that the extra thermal conductivity is caused by the
anomalous properties of carriers near the nodes---specifically,
that they acquire velocitiy parallel to the Fermi surface as well
as the conventional Fermi velocity. This problem will be the
subject of later publications. 
Our first
hypothesis for the $H$-field sensitivity was that spin-charge
separation was somehow involved in this large conductivity
enhancement and that $\kappa$ was being destroyed by the effects  of
Zeeman splitting, but 
this mechanism seems unable
energetically to account for the sharp $B$-dependence. The
mechanism is described below and involves
quasiparticles which are almost conventional.  

This problem
has been discussed by Lee and Simon\cite{3}, who derived scaling laws for
thermal conductivity which should be valid once the carriers are
confined to the nodes. This should be true for $T<T_c/2$, at
the very least. These scaling laws are
$$K_{xx}={\rm const} \times T\ F \ \Big ({\sqrt{B}\over T}\Big
)\eqno(1a)$$
and 
$$K_{xy}={\rm const} \times {T^2\over E_F}\ F' \ \Big
({\sqrt{B}\over T}\Big )\ .\eqno(1b)$$

These scaling laws are quite successful for $T<50^\circ$ for
optimally doped YBCO, and clearly the arguments of Ref (3) (which depend
on the result that energy levels in the nodes scale as $\sqrt{B}$, and hence
depend sensitively on magnetic field) contain the essence of the
phenomena. The purpose of this paper is to support their general
scaling argument with a specific picture of the nature of the
energy levels at the nodes in a magnetic field; and to predict
that broadened Landau-like levels will be observed, contrary to
previous discussions of these energy levels which ignore the
subtle properties of gap nodes.\cite{5}

One can discuss the physics semiclassically either as a Tomasch interference effect
involving Andreev scattering or as
a form of ``Landau level'' formation for excitations around the
``Dirac point'' nodes of the $d$-wave gap. The former point of
view is simpler. 
First we consider the effect of the uniform magnetic field and
will later discuss the effect of vortices. 
The $B$ field (along the $c$-axis,
perpendicular to the plane) causes a Larmor precession of all the
electrons around the Fermi surface according to
$$\hbar \dot k={e\over c}\ v_F\times B \ \ (\rm electrons\ and \
holes)
\eqno (2)$$
Here $e$ is the electron charge, negative in this case, $v_F$
the Fermi velocity at the nodal point, and $\hbar k$ the electron
momentum. Since we speak of an electron, $|k|$ may be assumed to
be somewhat $(\delta k)$ greater than $k_F$, giving an excess
kinetic energy $\delta E=v_F \delta k$. Holes obey the same
equation with the signs of both $e$ and $v_F$ changed, hence
with the 
$\rm\underline {same}$ direction of precession. 
The $k$-vector of an electron near a node is rotated away from the
node, and therefore  eventually encounters a gap equal to its excess kinetic
energy. We assume that in the superconducting state there are
point nodes in the gap, which in this region varies as 
$$\Delta={d\Delta\over dk}\ (k_\perp - k_{\rm node})= \hbar
v_\Delta\ (k_\perp-k_{node}) = \hbar v_\Delta\ \Delta k\ \ .\eqno(3)$$
where
$k_\perp$ is the momentum perpendicular to $v_F={\partial E\over
\hbar \partial k}$ and hence parallel to the Fermi surface, and
$v_\Delta$ is defined by (3). 
At this point the electron  must be Andreev reflected as a hole 
in the state of opposite momentum, supplying $(-2e)$ of charge to the condensate.
(Essentially, Fermion number as well as charge must be conserved.) This hole 
Larmor precesses in the 
same direction in $k$-space.

This at first appears to scotch the possibility of a periodic
process. However, it must be realized that the Andreev reflection
takes place not on a zero-momentum condensate but on a condensate
which macroscopically obeys London's equation
$$p ={eA\over c}$$
so that the momentum carried off by the pair is ${2eA\over c}$.
This means that the hole reappears not at the same momentum but
at a momentum shifted back by
$$\delta p=-2\hbar\ \Delta \,k$$
(it is easily verified that the change in ${2eA\over c} = 2\hbar \Delta\, k$.) 
Now the hole Larmor precesses from one barrier to the other, is
again Andreev reflected as an electron, and the process begins
again. 

The period of this process defines a frequency $f$ which will be
the excess energy level separation for electrons of that energy. It is
easy to estimate. For electrons of energy $\hbar \omega$, the
separation of the two barriers is 
$$\Delta k=2 \Bigg |{d k\over d\Delta}\Bigg |\cdot
\hbar\omega\eqno(4)$$
and the velocity in $k$-space gives 
$${dk_\perp\over dt}=f\cdot {\Delta k\over 2}={1\over 2}\ {ev_F\over \hbar
c}\cdot B$$
or,
$$ f={ev_F\over 4c\omega}\ \Bigg | {d\Delta\over dk}\Bigg |\ B$$
To  find the lowest energy state permitted by the uncertainty
principle , we set $f={\omega_0\over 2\pi}$ and obtain

\begin{eqnarray*} 
\hbar^2\omega^2_0 
&= {\pi\over 2}\ {\hbar e\over c}\ v_F\ B\ v_\Delta 
         = {\pi\over 2}\hbar^2({v_p\,v_\Delta\over \ell^2_H})  \\
&= {\pi\over 2}\ (\hbar \omega_c)\  E_F\cdot {v_\Delta  \over v_F}\\
&\hskip4truein(5)
\end{eqnarray*}

We have defined a velocity $v_\Delta ={d\Delta\over \hbar dk}$ to
represent the steepness of the energy gap at the nodes. This has
been estimated by Lee\cite{4} to be $\sim {1\over 8} v_F$. Using that
figure, and $E_F\sim 1{\rm ev}$, at 1T $h\omega_0\simeq 4.5 {\rm mev}$, 
which is of the right order of magnitude to explain the data. 

 A very physical way to think of this resonance is as a Landau
level in the Dirac points at each of the four nodes of the energy
gap, at which the quasiparticle excitations act like massless
Fermions of energy 
$$E^2_k=\hbar^2\,v_F^2\ (k_{||}-k_F)^2 +\Bigg ({d\Delta\over dk}\Bigg )^2
\ k^2_\perp\eqno(6)$$
These excitations change character as they precess around the
elliptical energy contour, from electrons through neutral
electron-hole mixtures to holes, but by the same argument given
above the quasiparticle precesses everywhere as though it had charge $e$ and a
velocity given by 
$$v_{Qp}={1\over \hbar}\ \nabla_k\ E_k\ \ \ , \eqno(7)$$ 
i.e, 
$$\hbar \dot k={e v_{Qp}\over c}\times B\ \ \ .\eqno(8)$$
As with the Andreev reflection picture, (8) seems a necessary
consequence of basic conservation laws. So long as the Larmor
frequency remains small compared with the maximum gap
$\Delta_0\sim 50 {\rm mev}$ 
tunneling through the gap maxima will play a minor role.  
(8) leads to the same result, (5),
for the cyclotron orbit frequency except for constants of order 1.  
Although in
momentum-space the orbit is slightly complicated by the momentum
shifts due to Andreev reflection on a moving condensate, these
shifts occur only parallel to the Fermi surface, and as far as
real space is concerned it is a conventional elliptical orbit. 

In order to go beyond the semiclassical picture, it is necessary
to solve the actual Bogoliubov-de Gennes equations for the
quasiparticle wave function near one of the $k$-space nodes of
the gap. These are
$${\cal H}\tilde{\psi}= h\omega\tilde{\psi}\eqno(9)$$
where $\tilde{\psi}$ is a two-component vector and ${\cal H}$ a
two-by-two tensor
$$
{\cal H}= 
\left( \begin{array}{cc}
v_F\cdot (p-{e\over c}A-mv_s)-\mu & \Delta^*                  \\
\Delta                        &(-v_F\cdot (p+{e\over c}A+mv_s)+\mu) 
\end{array}
\right)
\eqno(10)
$$
We will specialize to a single node of the gap and rotate  this
node to $k_x=k_F, \ k_y = 0$.
$$\Delta_0(k_x, \ k_y) = h k_y\cdot v_\Delta\ \ .$$
We have included the superfluid velocity in (10), thus
reintroducing the vortices. The superfluid velocity is equivalent
to a displacement of the entire Fermi surface (even near the
nodes) by $mv_s=p_s$ in momentum space, which leads to a
``doppler'' shift in the energy relative to the chemical
potential. Its sign agrees with that of $A$.

But it is also essential to include the position dependence of
$\Delta$ in the presence of a magnetic field, which is not
irrelevant because the off-diagonal, anomalous terms of (10)
represent processes in which a particle Andreev scatters into a hole,
with the charge of the missing pair going into the condensate. 
Such a process must also be momentum-conserving, so that if the
condensate has been accelerated by the magnetic field the
self-energy scattering must reflect that fact, and the pair
created in this process must carry the momentum of condensate
pairs.

What is the space-dependence of $\Delta$?  We can get a
reasonably good characterization in the ``London'' regime
$H_{c1}<< B<< H_{c2}$ where the vortices are dense enough to
overlap strongly so that the field is fairly uniform, while the
vortex cores still comprise a very small relative volume  $\propto
{B\over H_{c2}}$. Thus almost everywhere the field and current
satisfy London's equation as modified by the presence of a vortex
lattice. $\Delta$ is constant in magnitude, while its phase
$\varphi$ satisfies
$$j_s=n_sev_s={n_s\, e\over m}
(\hbar\nabla\varphi-{2e\over c}A)\ .\eqno(11)$$
$j_s$ is zero averaged over the interior of the sample, where the
field is nearly uniform. But near a vortex $j_s$ behaves like
$\vec \theta\over r$ and $\varphi=\theta$, the actual angle
measured around the vortex core. 
Note that $\Delta$ is single-valued
everywhere except at the vortex cores, though $\varphi$ of course
is not.

Now let us insert $\varphi$ into the Hamiltonian (10). There are
two ways to transform away the spatial dependence of $\Delta$: to
transform unitarily by
$$
U=\left( \begin{array}{cc}
e^{-i\varphi}&0\\
0& 1
\end{array}\right)
\qquad\qquad {\cal H}'=U^{-1}{\cal H} U
\eqno(12a)
$$
or by
$$
U=\left( \begin{array}{cc}
1 & 0\\
0&e^{i\varphi}
\end{array}\right)
\eqno(12b)
$$
(These are the only two single-valued transformations) In the
first case,
$${\cal H}_{holes}= 
\left( \begin{array}{cc}
v_F\cdot [(p+{e\over c}A)+mv_s]&\Delta_0\\
\Delta^*_0  &-v_F\cdot(p+{e\over c}A+mv_s)
\end{array}\right)
\eqno(13a)$$
using $\hbar\nabla\varphi=mv_s+{2e\over c}A$. 
This is the Hamiltonian appropriate for hole-like excitations,
since when $\Delta$ vanishes at the node, the wave-function is
purely hole-like. For electrons, we use (10b), and obtain
$$ ({\cal H})'_{electrons}= 
\left( \begin{array}{cc}
v_F\cdot(p-{e\over c}A-mv_s)&\Delta_0\\
\Delta_0&-v_F\cdot [(p-{e\over c}A)-mv_s]
\end{array}\right)
\eqno(13b)$$

If we neglect the 
$(p_{F}\cdot v_s)$ 
term, we come to the
conclusion that the basic form of the wave function is cyclotron
orbits in momentum space around Dirac points at the gap nodes,
with frequencies proportional to 
$\sqrt{B}$. 

The fact that these are two separate equations, one for electrons
and one for holes, reflects the fact that charge is actually
exactly conserved: when an electron is Andreev scattered into a
hole, the charge goes into the condensate and is recovered in the
inverse scattering. As Volovik has emphasized,\cite{5} gap nodes
are sites of chiral anomalies where supercurrent and
quasiparticle current are not separately conserved. 

But what of the 
$p_F\cdot v_s$ 
terms? These are quite strong in a
localized region around the vortex core. We predict that these
will have only a moderate effect.  The cyclotron orbits of low
order have wave-functions which occupy roughly an area of 
$n$
${\rm (magnetic \ lengths)^2}$, 
which is of order the size of the vortex lattice
unit cell, and it will cost energy to localize them further. I
suspect that the vortices act as moderately strong scattering
centers, broadening the Landau levels without changing them
quantitatively. (Since 
$\overline{v_s}=0$, 
these potentials will average to zero.)

Throughout this discussion, we have chosen a gauge such that
$A_y=0$, $A_x=By$. The dependence of $\Delta$ on $k_y$ is then
straightforward. But if we are to assume general gauge
covariance, clearly
$$\Delta (k_y)=\Delta \Big (k_{y}-{2e\ A_y\over \hbar c}\Big )$$
This in turn, implies that the $v_s$ term will also enter in the
$y$ direction by symmetry. Gathering all of those various strands
together, we come to the wave equation which nodal
particles (electrons, for example) obey,
$$E \tilde\psi={\cal H}\tilde\psi$$
$$ 
{\cal H}=
\left( \begin{array}{cc}
v_F\cdot[p-{e\over c}A-mv_s]&v_\Delta\cdot(p-{e\over
c}A-mv_s)\\
v_\Delta\cdot(p-{e\over c}A-mv_s)&-v_F\cdot(p-{e\over c}A-mv_s)
\end{array}\right)
\eqno(14)
$$
Thus the particles at nodal points are, unexpectedly, in all
respects equivalent to Dirac fermions in a vortex lattice and a
magnetic field.  At this point, the equations have become
essentially equivalent to those discussed by Volovik\cite{5}
in calculating Fermionic energy levels around the gap nodes in
the $He-3$ A phase. Volovik incorrectly asserts that the
equations for a superconductor in a magnetic field are not
equivalent to his equations, for which the effective magnetic
field, ``$B$''$=\nabla\times\hat{\ell}$, is provided by the order
parameter texture; but as we have shown above, this is not the
case near a gap node and the equations are identical except for
the doppler term $\vec v\cdot \vec P_s$, as is required by
conservation laws. 

In the absence of this term, the eigenvalues
are easily obtained by squaring the Hamiltonian.  Following
Volovik, we write
\begin{eqnarray*}
{\cal H} & = v_F\tau_x( {\hbar\over i}{\partial\over\partial x} - {e A_x\over c} )\\
         & + v_\Delta \tau_y({\hbar\over i}{\partial\over\partial y} - 
{e A_y\over c})
\end{eqnarray*}
where $\vec \tau$ is the Nambu ``isospin'' matrix homologous to
Pauli spin. Squaring this, we obtain 
\begin{eqnarray*}
{\cal H}^2 =& v^2_F({\hbar\over i}\,{\partial\over\partial x}
-{e A_x\over c})^2\\
+&v_\Delta^2({\hbar\over i}\,{\partial\over\partial_y}-{e A_y\over c})^2\\
+&{e\,\hbar\over c}v_f\,v_\Delta\,\tau_z\,B_z\\
&\hskip3truein {\rm (15)}\end{eqnarray*}

This leads to a spectrum of eigenvalues 
$$E^2_n=\hbar^2v_Fv_\Delta\Big ( {eB\over\hbar c}\Big )[(2n+1)|e|-\tau_z\,e)\eqno(16)$$
The striking feature of this is the ``zero mode'' $n=0$,
$\tau={e\over |e|}$, which is caused, basically, by the
topological effect of the vortex cores on the quasiparticles, and
in fact there is just one state per vortex for holes and one for
electrons.\cite{7}  

What do we predict, from this picture, for the thermal
conductivity? The physics is that of electrons in Landau orbits
whose spacing is proportional to $\sqrt{B}$, being scattered by
vortex cores (also spaced by $l_B\propto \sqrt{B}$), either
incoherently (at low fields, and high temperatures, where the
higher orbits are relevant and the lattice is disordered) or
possibly coherently, at high fields and low temperatures. In the
latter case there will be gaps in the spectrum and $\kappa$ could fall
to near zero: this is likely to be the cause of the plateaus
observed in some samples. 

The main predictions do not differ very much from those of Lee
and Simon's successful scaling laws, at least at low
temperatures.

As long as we can ignore the vortex cores the same conclusion
they reached, that the entire energy level structure scales as
$\sqrt{B}$, seems entirely valid, so that the scaling function
will be
$F(\sqrt{B}/T)$. (Just by dimensional analysis: changing B is
equivalent to a shift in $p$ by a factor $l^{-1}_B$.) The
justification given here is slightly more rigorous than that 
in the reference. The function which Ong and Krishana use for $K_{xx}$,
$$F={1\over 1+p\,B}$$
obeys the Lee-Simon scaling law over most of the range where it
should, $p$ having roughly the form $1/T^2$; and as to order of
magnitude, $p B$ is of order unity where the lowest cyclotron
orbit $\omega_0$ is of order $T$. 

At low fields and moderate temperatures, we could imagine that
the classical formula
$$\kappa\propto T\sigma \propto T{1\over 1 +\omega^2_c\tau^2}\eqno(17)$$
would be usable. $\omega^2_c$ is proportional to $B$, and it is
reasonable that $1/\tau$ caused by scattering on the cores would
be independent of $B$ and simply proportional to the density of final 
states $\propto T$
(though by no means obvious.)
At high fields the states become increasingly localized and
eventually only the zero mode is occupied. It is not at all clear
why $\sigma\alpha {1\over B}$ in this region and in fact
plateaus with $\sigma\simeq 0$ are occasionally observed. 
(There are actually $\underline{\rm more}$ low-energy states, not
fewer; but they seem likely to localize better as $B$ increases.) 
The experiments show us that
the vortex current scattering effect is, as expected, similar in
magnitude to the cyclotron frequency itself, so that the
cyclotron orbits are not sharp in energy except where the
scattering is coherent.

As for $\kappa_{xy}$, as pointed out by Lee and Simon this is smaller
than $\kappa_{xx}$ by a factor $\sim T/E_F$ because it comes from
hole-particle dissymmetry. It is significant that over a range 
of intermediate $B$'s it fits fairly well to $\sqrt{B}/1+pB$,
(see Fig. (2),) which would follow from semiclassical arguments
from the Landau orbits.
But since this is in principle outside the ``node''
approximation, it will not be discussed in detail here. 

The ``spin gap'' phase seems to be one in which the mode
structure of the pseudogap is already developed. (This is the
picture proposed by Lee and Nagaosa\cite{4}, following Baskaran
and Zou, and borrowed by Fisher, Balents and Nayak\cite{6}.) Thus
some of the above considerations may be relevant for critical
phenomena near $T_c$, in the spin gap phase.

\bigskip

I would like to acknowledge very extensive discussions of data
and theory with N.P.~Ong, as well as 
the cooperation of K.~Krishana in making his extensive data readily available.
Discussions with C.~Nayak, R.B.~Laughlin, and F.D.M.~Haldane were
very stimulating.

\vfill\eject

\end{document}